\documentclass[prb,twocolumn,showpacs,superscriptaddress]{revtex4}

\usepackage{bm}
\usepackage{graphicx}

\def \WHOZA{Institute of Experimental Physics, University of Warsaw, ul. Ho\.za 69, 00-681 Warszawa, Poland}
\def \WPAC{Institute of Physics, Polish Academy of Sciences, al. Lotnik\'ow 32/46, 02-688 Warszawa, Poland}

\begin{document}
\preprint{Aug 07, 2014}

\title{Coherent precession of an individual 5/2 spin}

\author{M. \surname{Goryca}}\email{Mateusz.Goryca@fuw.edu.pl}\affiliation{\WHOZA}
\author{M. \surname{Koperski}}\affiliation{\WHOZA}
\author{P. \surname{Wojnar}}\affiliation{\WPAC}
\author{T. \surname{Smole\'nski}}\affiliation{\WHOZA}
\author{T. \surname{Kazimierczuk}}\affiliation{\WHOZA}
\author{A. \surname{Golnik}}\affiliation{\WHOZA}
\author{P. \surname{Kossacki}}\affiliation{\WHOZA}

\date{\today}

\begin{abstract}
We present a direct observation of a coherent spin precession of an individual Mn$^{2+}$ ion, having both electronic and nuclear spins equal to 5/2, embedded in a CdTe quantum dot and placed in magnetic field. The spin state evolution is probed in a time-resolved pump-probe measurement of absorption of the single dot. The experiment reveals subtle details of the large-spin coherent dynamics, such as non-sinusoidal evolution of states occupation, and beatings caused by the strain-induced differences in energy levels separation. Sensitivity of the large-spin impurity on the crystal strain opens the possibility of using it as a local strain probe.\end{abstract}


\maketitle

Coherent behavior of an individual spin in a solid has been demonstrated in several two- and three-level (so-called \emph{lambda}) systems. Among them are systems such as a single electron \cite{Loss_1998_PRA, Koppens_2006_Nature, Atature_2006_Science, Press_2008_Nature, Berezovsky_2008_Science}, a hole \cite{Brunner_2009_Science} or an exciton \cite{Stievater_2001_PRL,Kamada_2001_PRL} in a quantum dot (QD), a nitrogen-vacancy center in diamond \cite{Jelezko_2004_PRL,Gaebel_2006_Nature_Phys} or a rare-earth ion in ceramic crystal \cite{Siyushev_2014_Nature_Comm}. However, coherence in more complex multi-level large-spin systems was, up to now, inaccessible experimentally. A possibility to change this situation emerged due to a constant progress in precise control of the doping of semiconductors. Within the last few years it opened the whole new field of solotronics (solitary dopant optoelectronics) \cite{Koenraad_2011_Nature_Mat}, focused on properties of individual ions and defects embedded in a semiconductor lattice. In particular, large-spin magnetic impurities have attracted a lot of intrest. They offer desirable features of solitary spins, such as strong localization leading to weak coupling with host crystal. Many experiments performed on different systems \cite{Besombes_2004_PRL, Kudelski_2007_PRL, Kobak_2014_Nature_Comm} provided substantial knowledge on the physics of these objects, including the demonstration of optical readout and manipulation of the electronic spin state \cite{LeGall_2009_PRL, Goryca_2009_PRL2, Goryca_2010_Phys_E}. However, those experiments concerned only non-coherent phenomena. Coherent measurements, including observation of Rabi oscillations and a long coherence time, were performed only on ensembles of magnetic ions, e.g., colloidal QDs containing magnetic dopants \cite{Ochsenbein_2011_Nature_Nano}, transition metal ions embedded in ZnO or MgO crystal \cite{Tribollet_2008_EPL, Bertaina_2009_PRL, George_2013_PRL}, or ensembles of molecular magnets \cite{Ardavan_2007_PRL, Bertaina_2008_Nature}. Up to now, the coherent dynamics of an individual large-spin particle has been considered only theoretically, showing e.g., possibility of using a molecular magnet for quantum information storage and processing \cite{Leuenberger_2001_Nature, Bogani_2008_Nature_Mat}, or a magnetic ion as a multi-qubit system \cite{Gun_2013_QIP} with optical interface for readout and manipulation \cite{Reiter_2009_PRL}. 

In the present letter we show a direct observation of a coherent spin precession of an individual Mn$^{2+}$ ion, having both electronic and nuclear spins equal to 5/2, placed in magnetic field applied perpendicularly to the quantization axis of the system (see below). The idea of the experiment is to probe the spin state of the single Mn$^{2+}$ ion in a time-resolved measurement of the absorption of a QD containing such impurity. The QD is resonantly excited with two consecutive laser pulses, while the absorption is analyzed versus the relative delay between them. 

The sample studied in the experiment contains a single layer of self-assembled CdTe/ZnTe QDs. The dots contain a low amount of Mn$^{2+}$ ions, so that selection of single dots with exactly one magnetic ion is possible with the use of a micro-photoluminescence setup. The emission of the QDs is excited either continuously with a rhodamine 6G dye laser emitting in the range 570-610 nm or in a pulsed regime with an optical parametric oscillator (OPO) tuned to the same range. The pulses generated with OPO are spectrally narrowed and temporally broadned with the use of etalon. Its thickness determines the spectral and temporal width to about 0.6~meV and 1~ps, respectively. The OPO beam is split into two beams, one of which is passed through a mechanical delay line enabling a precise control over the delay between the two pulses arriving at the sample. 

\begin{figure}[ht]
\begin{center}
\includegraphics[width=80mm]{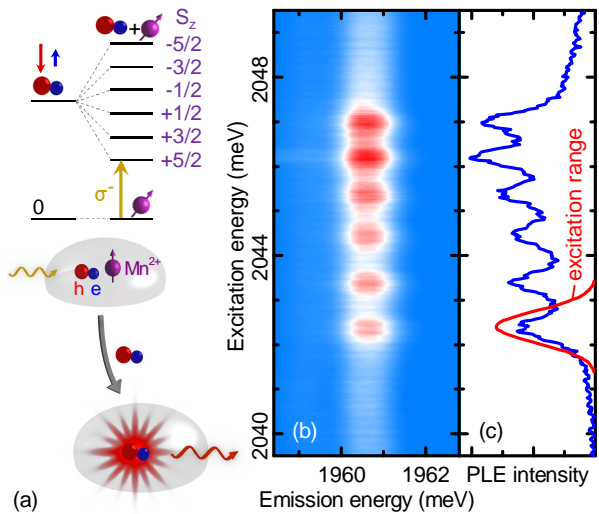}
\end{center}
\caption[]{(Color online) (a) A schematic diagram of the excitation process of two coupled QDs with a single Mn$^{2+}$ ion in the absorbing one together with a diagram of excitonic energy levels in a QD without and with such ion, for $I_z$~=~-1 polarized exciton. Each optical transition in the dot with Mn$^{2+}$ is related to a specific spin state of the ion. The arrow indicates the transition chosen for most of the experiments. (b) Photoluminescence excitation map (density plot of PL intensity versus emission/excitation photon energy) of the two coupled QDs. (c) Photoluminescence excitation spectrum of the QD with a single Mn$^{2+}$ ion together with the spectrum of the pulsed laser used in time-resolved measurements.}
\label{map}
\end{figure}

The QD containing a single magnetic impurity is selected from the self-assembled system with the use of photoluminescence-excitation (PLE) measurements. The QD absorption is detected by using the excitation transfer to a neighboring dot from which the luminescence is recorded \cite{Kazimierczuk_2009_PRB,Goryca_2009_PRL2,Goryca_2010_PRB,Koperski_2014_PRB} (see Fig.~\ref{map}(a)). Excitons created resonantly in the dot containing the single magnetic ion do not recombine in this dot, as the transfer time to the coupled one is of the order of a few ps \cite{Kazimierczuk_2009_PRB}, much shorter than their radiative lifetime. The recombination takes place in the coupled QD with a lower exciton ground energy, resulting in the emission of a photon at energy lower by about 100-200 meV than the excitation energy. The luminescence intensity of the emitting QD is monitored either at the neutral or charged exciton line, depending on the mean charge state of that dot. Such approach assures a very short exciton lifetime in the absorbing QD, crucial for the temporal resolution of the experiment. A fingerprint of the presence of the single Mn$^{2+}$ ion in the absorbing dot originates from the exciton-Mn$^{2+}$ exchange interaction. Since the Mn$^{2+}$ electronic spin is equal to 5/2, this interaction leads to a six-fold splitting of the excitonic ground state \cite{Besombes_2004_PRL}. As a result, six excitonic lines are visible in the absorption spectrum of the QD, which corresponds to six well separated bright spots on the PLE map, as shown in Fig.~\ref{map}(b). In the time-resolved experiment, the pulsed laser is tuned to one of them. That leads to the formation of an exciton-Mn$^{2+}$ complex with specifically chosen Mn$^{2+}$ spin projection onto the growth axis ($z$), which defines the exciton quantization axis. - see Fig.~\ref{map}(a).

The measurements of the Mn$^{2+}$ spin dynamics in magnetic field are performed in a degenerate pump-probe regime, with both laser pulses exciting the transition corresponding to the same Mn$^{2+}$ spin state. When the photon from the first pulse is absorbed by the QD, there is a little probability that during the lifetime of the exciton a scattering process takes place, leading to a change of the Mn$^{2+}$ state. Such a process can involve for example an exciton-Mn$^{2+}$ spin flip-flop. As a result, after the pump pulse the Mn$^{2+}$ spin state corresponding to the pumped transition is depleted as compared to the situation before the pulse. A similar process under continuous-wave excitation was already investigated and described in Ref. \onlinecite{LeGall_2011_PRL}.

The second pulse probes the spin state of the Mn$^{2+}$ ion. The probability of absorption is proportional to the probability of finding the ion in the state defined by the energy and polarization of the laser light. By measuring the absorption while varying the delay between the two pulses we are able to probe the evolution of the selected Mn$^{2+}$ spin state occupation. As the system under investigation is placed in a magnetic field perpendicular to the quantization axis and is not in a relaxed state, one should expect to observe oscillation of the occupation of the state initially depleted by the pump pulse. The frequency of the oscillation is given by the magnetic field and Land\'e factor of the Mn$^{2+}$ ion.

In order to have a more complete insight into complex spin evolution of the large-spin Mn$^{2+}$ ion the experiment is performed also in a non-degenerate regime. The pump and probe pulses are resonant with transitions corresponding to different spin states of the Mn$^{2+}$ ion. In principle this can be realized either by the use of different energies for the two pulses, or the same energies but different circular polarizations. In our experimental setup the latter approach is used.

\begin{figure*}[htp]
\begin{center}
\includegraphics[width=160mm]{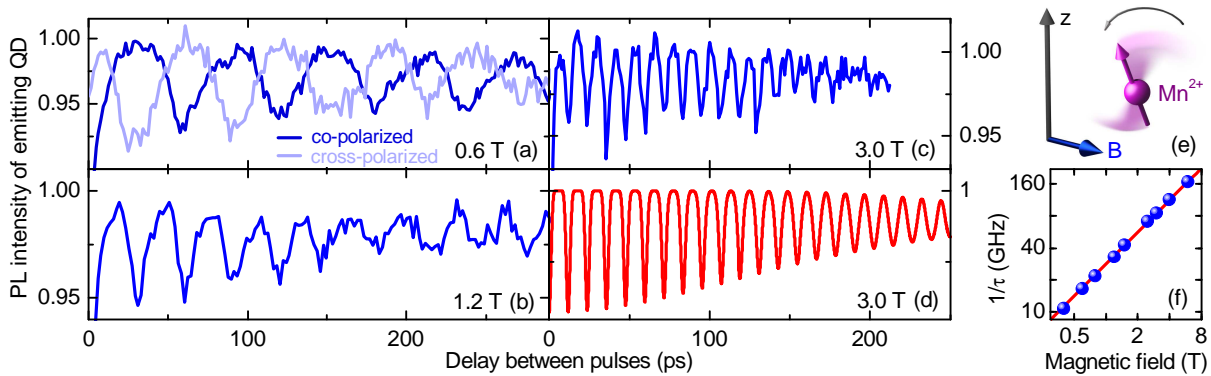}
\end{center}
\caption[]{(Color online) (a-c) Measured and (d) calculated evolution of the occupation of the $S_z$ = +5/2 state of the Mn$^{2+}$ ion embedded in a QD for indicated values of magnetic field applied in Voigt configuration (perpendicularly to the quantization axis of the system). At magnetic field equal to 0.6 T signal is measured both for co- and cross-polarized pump and probe pulses. (e) A schematic diagram of Larmor precession of the Mn$^{2+}$ spin in the magnetic field ($B$) applied perpendicularly to the quantization axis ($z$), given by the anisotropy of the exciton confined in the QD. (f) Dependence of the measured Larmor precession frequency on the magnetic field (points) together with linear fit (solid line).}
\label{osc}
\end{figure*}

The pump-probe experiment is performed for several pairs of coupled QDs. Representative results are shown in Fig. \ref{osc}(a-c). The series of graphs show the luminescence intensity of emitting QD, coupled to the one containing the single Mn$^{2+}$ ion, measured as a function of delay between the two laser pulses. The signal measured upon application of low magnetic field (B=0.6~T) is presented in Fig. \ref{osc}(a). The two curves represent data measured in a degenerate and a non-degenerate regime. In the first case both pump and probe pulses correspond to the $S_z$ = +5/2 state of the Mn$^{2+}$ ion. In the second one the pump pulse is adjusted to the $S_z$ = +5/2 state, while the probe pulse to the $S_z$ = -5/2 state. Oscillations of the measured signal are clearly visible, indicating the coherent Larmor precession of the individual magnetic ion. The two curves present the same frequency of the oscillations, but an opposite phase, as expected for measurements in degenerate and non-degenerate regime.

Experimental data obtained for higher values of magnetic field are presented in Fig. \ref{osc}(b-c). The frequency of the oscillations increases with the magnetic field and, as shown in the Fig. \ref{osc}(f), follows the linear dependence on the field. The Land\'e factor of the Mn$^{2+}$ ion deduced from this dependence is equal to 2.0. The amplitude of the oscillations clearly decreases for longer delays, with characteristic time of about 200~ps and with no evidence of magnetic field dependence.

\begin{figure}[ht]
\begin{center}
\includegraphics[width=80mm]{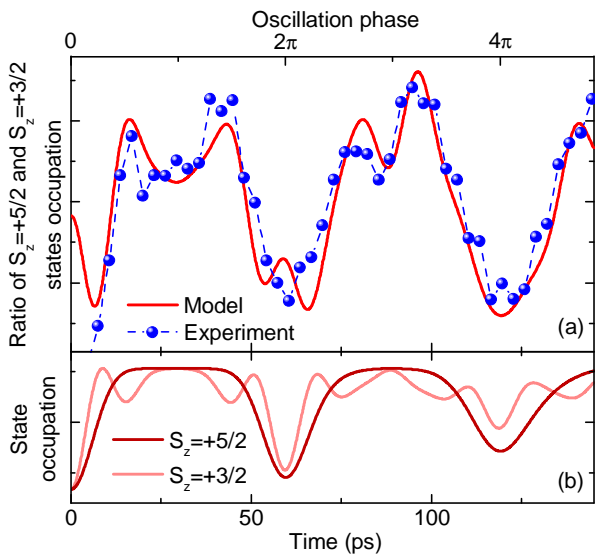}
\end{center}
\caption[]{(Color online) (a) Measured (points) and calculated (solid line) curve representing the time-dependent ratio of the occupation of the $S_z$ = +5/2 and $S_z$ = +3/2 states of the Mn$^{2+}$ ion in a degenerate pump-probe measurements at $B$ = 0.6~T. The theoretical curve obtained with the model described in the text (no fitting parameters except for vertical normalization) is convoluted with the Gaussian curve of the width equal to 3~ps, related to the experimental resolution. (b) Theoretical curves representing the occupation evolution of the $S_z$ = +5/2 and $S_z$ = +3/2 states before convolution with the Gaussian curve.}
\label{diff}
\end{figure}

The complex properties of spin 5/2 are revealed when analysing in detail the occupation evolution of its spin states. It is not described with the simple sine curve. In particular, measured curves expose the subtle differences between the occupation evolution of different states of the magnetic impurity. An example result presenting such differences is shown in Fig. \ref{diff}(a). It compares two experimental curves taken with both pump and probe pulses tuned to the transitions related to the $S_z$ = +5/2, or $S_z$ = +3/2 states.

Experimental data representing the Mn$^{2+}$ spin evolution can be quantitatively described taking into account the Zeeman effect for the electronic spin, the hyperfine coupling, the cubic crystal field, and the biaxial strain, which is usually a dominant strain component in CdTe QDs grown on ZnTe. This gives following Hamiltonian \cite{Quazzaz_1995_SSC,Goryca_2009_PRL1}: 

\begin{eqnarray}\label{eq:eq1}
\mathcal{H}=g\mu_B\overrightarrow{B}.\overrightarrow{S}+A\overrightarrow{I}.\overrightarrow{S}
+D_{0}[S_{z}^{2}-\frac{S(S+1)}{3}] \nonumber
\\
+\frac{a}{6}[S_{x}^{4}+S_{y}^{4}+S_{z}^{4}-\frac{S(S+1)(3S^{2}+3S-1)}{5}]
\end{eqnarray}

where $g=~$2.0 is Mn$^{2+}$ Land\'e factor, $A=~$680~neV is the hyperfine coupling constant, $a=~$320~neV the cubic crystal field splitting, and $D_0$ describes the effect of the biaxial strain component. Note that all above parameters are precisely known from EPR measurements of Mn$^{2+}$ ions in bulk crystals \cite{Quazzaz_1995_SSC}, except for $D_0$ as it depends on the properties of a particular QD. Importantly, the term related to strain only slightly affects the precession frequency of the Mn$^{2+}$ ion \cite{Quazzaz_1995_SSC}. Therefore, its influence on the observed signal in a short time scale (of the order of 100~ps) can be neglected. In order to obtain the theoretical curve representing the occupation evolution of a selected spin state we take the initially thermalized Mn$^{2+}$ ion, deplete the state corresponding to the pump pulse energy at $t$ = 0 and calculate the subsequent evolution of the ion spin state using the density matrix formalism. Additionally, for a comparison with the experimental data, such evolution is convoluted with the Gaussian curve of the width equal to 3~ps, related to the overall experimental resolution.

As shown in the Fig. \ref{diff}(a), the difference between the two signals related to the $S_z$ = +5/2 and $S_z$ = +3/2 states is clearly reproduced with the theoretical curve. Note that, as we consider roughly the first 100~ps of the system evolution, the theoretical curve is computed with no fitting parameters except for vertical normalization. 

\begin{figure}[ht]
\begin{center}
\includegraphics[width=80mm]{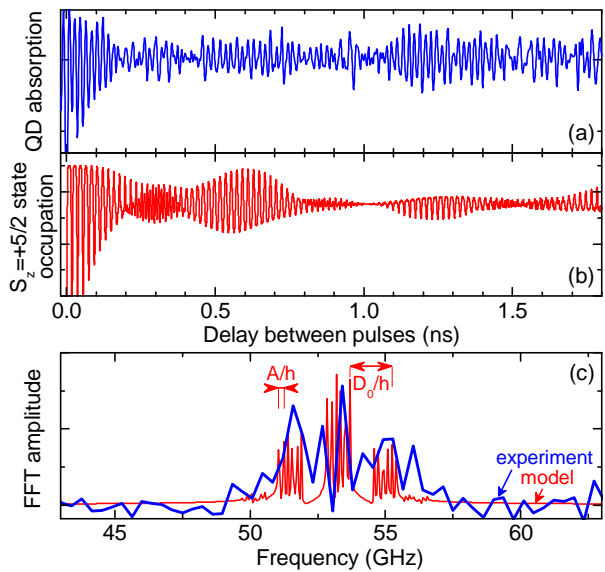}
\end{center}
\caption[]{(Color online) Measured (a) and calculated (b) curve corresponding to the occupation evolution of the $S_z$=+5/2 state of the Mn$^{2+}$ ion for large values of the delay between the pump and probe pulses, together with the Fourier transform of presented signals (c).}
\label{beat}
\end{figure}

The decrease of the oscillations amplitude for the time scale longer than 100~ps, visible in Fig. \ref{osc}(a-c) may, in principle, originate from two factors: the Mn$^{2+}$ spin decoherence or the influence of interactions additional to the Zeeman effect, making the whole process more complicated than the simple Larmor precession. The latter case is more desirable not only from the applicational point of view, but also opens the possibility of experimentally exploiting the Mn$^{2+}$ coherence in a long timescale. Here we show that the observed Mn$^{2+}$ dynamics can be fully described without taking into account any coherence loss. An example fit to the experimental curve shown in Fig. \ref{osc}(c) is presented in Fig. \ref{osc}(d). It is obtained with the model taking into account all Hamiltonian terms described above. Two of them lead to the observed decrease of the oscillations amplitude: the interaction of the Mn$^{2+}$ electronic spin with the nuclear one and the term describing strain-induced component of the crystal field. As the total nuclear spin is equal to 5/2, there are - as in the case of the electronic spin - six possible projections onto the quantization axis. The hyperfine interaction between the two spins acts on the electronic one as a small effective magnetic field. Therefore the precession frequency of the electronic spin slightly depends on the nuclear spin orientation. On the other hand, the strain-induced component of the crystal field changes the energy separation between different pairs of the Mn$^{2+}$ spin states \cite{Quazzaz_1995_SSC}. Thus it introduces additional components to the frequency spectrum of the Mn$^{2+}$ precession. The relative importance of these two effects depends on the ratio between the $A$ and $D_0$ parameters of the Hamiltonian. Both of them also lead to the presence of beatings in the observed evolution of the Mn$^{2+}$ spin. Such beatings are indeed observed for delays between the pump and probe pulses as long as 1.8~ns, which is shown in Fig. \ref{beat}(a). The Fourier transform of the observed signal reveals a broad structure centered at the frequency determined by the magnetic field (53.2~GHz at 1.9~T) of the width equal to about 4~GHz, clearly visible in Fig. \ref{beat}(c). Such width corresponds to the energy splitting an order of magnitude larger than $A$. This indicates that the observed beatings are caused mainly by the crystal field, which is also the main limiting factor for the characteristic time of the initial decrease of the observed oscillations amplitude. The best fit to the experimental data, shown in Fig. \ref{beat}(b), is obtained using the parameter describing the strain component $D_{0}=~$6.5~$\mu$eV.

Due to experimental limitations, delays longer than 1.8~ns are not investigated. Nevertheless, the presence of coherence of the Mn$^{2+}$ electronic spin after 1.8~ns permits us to regard this value as a lower bound of the $T_{2}$ time. The upper bound is given by the $2\cdot T_{1}$ measured previously on the same system, being of the order of a millisecond \cite{Goryca_2009_PRL2}. 

To conclude, using the pump-probe technique to perform a time-resolved measurement of the absorption of a single QD containing a single Mn$^{2+}$ ion, we have directly shown a coherent precession of an individual 5/2 spin placed in magnetic field. The multitude of the large-spin ion energy levels entails new effects when compared to simple two-level systems. Those effects include e.g., non-sinusoidal evolution of states occupation, and beatings caused by the strain-induced differences in energy levels separation. The latter effect opens the possibility of tailoring the vicinity of the magnetic ion to control its coherent evolution or, on the other hand, using the large-spin impurity as a probe of a local strain.

\textbf{Acknowledgments:}
This work was supported by the Polish National Science Center under decisions 
DEC-2011/02/A/ST3/00131, DEC-2012/07/N/ST3/03130, DEC-2012/07/N/ST3/03665, DEC-2013/09/B/ST3/02603 and DEC-2013/09/D/ST3/03768, by the Polish Ministry of Science and Higher Education in years 2012-2016
as research grant "Diamentowy Grant", and by the Foundation for Polish Science (MISTRZ programme). Project was carried out with the use of CePT, CeZaMat, and NLTK infrastructures financed by the European Union - the European Regional Development Fund within the Operational Programme "Innovative economy" for 2007 - 2013. 

We thank Sophia Economou and \L{}ukasz Cywi\'nski for fruitful discussion as well as Aleksander Bogucki, Wojciech Pacuski, \L{}ukasz K\l{}opotowski, Micha\l{} Nawrocki and Jan Suffczy\'nski for help in manuscript preparation.

\end{document}